\documentstyle[preprint,aps]{revtex}
\draft
\begin{document}
\title{Isospin non-equilibrium in heavy-ion collisions at intermediate
energies}
\bigskip
\author{\bf Bao-An Li$^{a}$ and Sherry J. Yennello$^{b}$}
\address{a Cyclotron Institute and Department of Physics\\
b Cyclotron Institute and Department of Chemistry\\
Texas A\&M University, College Station, TX 77843, USA}
\maketitle

\begin{quote}
We study the equilibration of isospin degree of freedom in
intermediate energy heavy-ion collisions using an isospin-dependent
BUU model. It is found that there exists a transition from the isospin
equilibration at low energies to non-equilibration at high energies
as the beam energy varies across the Fermi energy in central,
asymmetric heavy-ion collisions. At beam energies around 55 MeV/nucleon,
the composite system in thermal equilibrium but isospin non-equilibrium
breaks up into two primary hot residues with N/Z ratios closely related
to those of the target and projectile respectively. The decay of
these forward-backward moving residues results in the strong
isospin asymmetry in space and the dependence of the isotopic
composition of fragments on the N/Z ratios of the target and projectile.
These features are in good agreement with those found recently in
experiments at NSCL/MSU and TAMU, implications of these findings
are discussed.
\end{quote}

\newpage
With the availability of high intensity radioactive beams at several existing
and proposed facilities, the isospin degree of freedom in nuclear
reactions can be studied in large domains of beam energies and
projectile-target combinations. In recent experimental studies of
the isotopic composition of intermediate mass fragments from
central collisions of $^{40}Cl,~ ^{40}Ar$
and $^{40}Ca$ with $^{58}Fe$ and $^{58}Ni$ at $E_{beam}/A$=25, 35, 45
and 53 MeV at NSCL/MSU and TAMU\cite{sherry1,sherry2,sherry3}, it was
found that at $E_{beam}/A$=25 and 35 MeV the isotopic
ratios $^9Be/^7Be, ~^{11}B/^{10}B$ and $^{13}C/^{12}C$ increase
linearly as a function of the increasing $(N/Z)_{cs}$ ratio of the combined
target and projectile system, but are independent of the N/Z ratio
of the target or projectile.
This observation strongly indicates the establishment of isospin
equilibration in composite systems formed in these reactions
before the emission of fragments.
In fact, previous studies on similar systems at a much lower energy showed
that the isospin degree of freedom was one of the fastest to
equilibrate\cite{gatty}. The most striking and unexpected feature was
observed from the isotopic ratios in central collisions at
$E_{beam}/A=$ 53 MeV. It was found that the isotopic ratios depend
on the N/Z of the target and projectile for entrance
target-projectile combinations having the same $(N/Z)_{CS}$ ratio,
such as $^{40}Ca+^{58}Fe$ and $^{40}Ar+^{58}Ni$.
Moreover, data at very forward and backward angles
also show isotope ratios that are not simply a function of the $(N/Z)_{cs}$.
It was found that light fragments at backward angles have a much greater
dependence on $(N/Z)_{target}$, while at forward angles have a much greater
dependence on $(N/Z)_{projectile}$, and could not be explained simply by the
pre-equilibrium emission. These features clearly demonstrate that
the isospin degree of freedom is not equilibrated in the reactions at
$E_{beam}/A=$ 53 MeV on the time scale of emission
of the fragments.

The above observation has some deep implications on the reaction mechanism
leading to multifragmentation. It not only establishes a relative time scale
of multifragmentation in these reactions but also indicates that
the assumption of isospin equilibrium taken for granted in various
statistical models for nuclear multifragmentation at intermediate
energies must be improved at least.
To our best knowledge, the prominent features observed in these
experiments were neither predicted nor observed previously in any model
simulations. On the contrary,
some models which are able to explain many other aspects of
multifragmentation and reaction dynamics of intermediate energy
heavy-ion collisions are unable to explain these features.
In particular, statistical models based on the assumption of isospin
and thermal equilibrium of the composite system fail to show
any entrance channel effect\cite{sherry1} as one would expect.
Calculations using an intranuclear cascade code ISABEL\cite{yariv}
show that the N/Z of the residue is very close to the N/Z of the
initial combined system\cite{sherry1} and thus also failed to
reproduce the features observed at $E_{beam}/A=$ 53 MeV.
Although the exact origin of this failure is not so clear to us,
one expects that the reaction dynamics at these energies should go beyond the
nucleon-nucleon cascade and invoke the nuclear mean field.
Therefore, the experimentally observed transition from isospin equilibration
to non-equilibration as the beam energy increases from below to above
the Fermi energy remains an interesting question for theoretical
studies.

In this Letter we report on results of a study on the
equilibration of isospin degree of freedom using an
isospin-dependent Boltzmann-Uehling-Uhlenbeck (BUU) transport
model for heavy-ion collisions at intermediate
energies\cite{bertsch88,bauer}.
The isospin dependence comes into the model\cite{li}
through both the elementary nucleon-nucleon
cross sections $\sigma_{12}$ and the nuclear
mean field $U$. Here we use the experimental nucleon-nucleon
cross sections with the explicit isospin dependence\cite{data}.
The isospin dependence resides in the fact that
the cross section of neutron-proton collisions
is about three times that of neutron-neutron or
proton-proton collisions at energies interested here.

The nuclear mean field $U$ including the isospin symmetry
term is parameterized as
\begin{equation}
      U(\rho,\tau_{z}) = a (\rho/\rho_0) + b (\rho/\rho_0)^{\sigma}\
	+(1-\tau_{z})V_{c}+C\frac{\rho_{n}-\rho_{p}}
	{\rho_{0}}\tau_{z}.
\end{equation}
Here $\rho_{0}$ is the normal nuclear matter density,
$\rho$, $\rho_{n}$ and $\rho_{p}$ are the nucleon, neutron and proton
densities, respectively. $\tau_{z}$ equals 1 or -1 for neutrons or protons,
respectively, and $V_{c}$ is the Coulomb potential. Corresponding to other,
probably more complete forms of Skyrme forces, other forms of
parameterization for the mean field are possible\cite{remaud,farine,sobotka}.
Here we use the so called soft nuclear equation of state with the
nuclear compressibility $K=200$ MeV.
The symmetry term comes from averaging over the constituent
two-body forces with Heisenberg components proportional to
$(\vec{\tau}_{i}\cdot\vec{\tau}_{j})$\cite{hodgson}. Its strength
C can be deduced from experiments (e.g. nuclear symmetry energies, optical
potentials for nucleon scatterings, excitations of analogue states in (p,n)
reactions). However, the strength deduced varies significantly from
reaction to reaction and also depends on the energy of the
nucleon\cite{hodgson}. The parameter C
used in transport model calculations also scatters within a large
range\cite{farine,betty,pawel}. In the present study, we
use a constant $C=32$ MeV.

First, to study the energy dependence of the reaction mechanism we show
in Fig.\ 1 the density contours in the reaction plane at t= 200 fm/c
in head-on collisions of Ar+Ni at $E_{beam}/A=$ 25, 35, 45 and 55 MeV.
The solid contours with $\rho=\rho_0/8$ essentially bound the
composite systems or heavy residues formed in the reactions. The
dotted contours with $\rho=0.05 \rho_0$ are the representatives of
free nucleons mainly from pre-thermal-equilibrium emissions.
The most interesting feature in Fig.\ 1 is the formation of two
heavy residues in the reaction at $E_{beam}/A=55$ MeV. While by following
the evolution of the heavy residues until 300 fm/c, we found no
breakup happening to the residues formed in the reactions at
$E_{beam}/A=$ 25, 35 and 45 MeV.

Thermal equilibrium of the heavy residues can be examined by studying the
quadrupole moment $Q_{ZZ}$ defined as
\begin{equation}
Q_{zz}(t)=\int\frac{d\vec{r}d\vec{p}}{(2\pi)^3}(2p^2_z-p^2_x-p^2_y)
f(\vec{r},\vec{p},t).
\end{equation}
Where $f(\vec{r},\vec{p},t)$ is the Wigner function from the BUU model
calculations. It is clear that $Q_{zz}=0$ is a necessary but not a
sufficient condition for thermal equilibrium. We found that the
quadrupole moments of heavy residues
formed in the reactions at $E_{beam}/A=$ 25, 35 and 45 MeV have been damped to
about zero by the time of about 300 fm/c. Since the heavy residue formed in the
reaction at $E_{beam}/A=$ 55 has already broken up into two pieces
at about 200 fm/c, it is interesting to examine more closely the
evolution of $Q_{zz}$ in this reaction. This is done in Fig.\ 2
where $Q_{zz}$ of the heavy residue
scaled by its mass number $A_{res}$ is shown as a function of time.
For comparison we also show in this figure the evolution of
$Q_{zz}$ for the reaction at $E_{beam}/A=$ 45. It is seen that
the residues are very close to thermal equilibrium at about 200 fm/c
indicated by the very small, decreasing amplitude of the
oscillation around zero. In the case of $E_{beam}/A=55$ MeV, the
oscillation is shifted towards a small positive value
indicating a slight longitudinal collectivity or transparency.
This finding is actually in very good agreement
with the transparency found recently at Ganil for the same reaction system at
similar beam energies using the INDRA detector\cite{borderie}.

Are the heavy residues observed above in isospin equilibrium ?
To answer this question
we show in Fig.\ 3 the numbers of neutrons and protons in the residues
on the left ($Z < 0$) and right ($Z\geq 0$) side of
the origin in the head-on reactions of Ar+Ni at $E_{beam}/A=$25, 35, 45
and 55 MeV. The solid lines are the numbers of protons from the
projectile, while the dotdashed lines are the numbers of neutrons from
the projectile which comes in originally from the left.
The dashed lines are the numbers of protons from
the target, while the dotted lines are the numbers of neutrons from
the target which comes in originally from the right
in the center of mass frame. It is seen that the numbers
of neutrons and protons on both sides undergo damped oscillations.
The damping is mainly due to nucleon-nucleon collisions and particle emissions.
While the oscillations are due to both the restoring force of the
mean field and nucleon-nucleon collisions.
At $E_{beam}/A=$ 25 MeV the neutron and proton numbers on the two sides
have become very close to each other and the amplitude of oscillation
is becoming rather small by the time of 300 fm/c. It indicates
that the heavy residue is very close
to the isospin equilibrium characterized by the space-time independent proton
and neutron number distributions. At $E_{beam}/A=35$ MeV the oscillation is
damped at a higher rate to a distribution of particles
close to the isospin equilibrium. The isotopic composition of fragments
emitted from the residues of these low energy reactions
after about 300 fm/c would therefore essentially
reflect the $(N/Z)_{cs}$ of the initial composite system, and there is little
forward-backward asymmetry. These features are in good agreement with
those found in the preliminary data of $E_{beam}/A=$25 and 35
MeV\cite{sherry2,sherry3}.

At higher energies, such as $E_{beam}/A=$ 45 and 55 MeV,
there is little oscillation across the overlapping region between the
target and projectile. This is mainly because
the incoming momenta of projectile-nucleons and target-nucleons
are too high for the mean field and nucleon-nucleon collisions
to be effective enough during a relatively
short reaction time to reverse the directions of motion of many
nucleons. As a result, there exists a large isospin asymmetry or
non-equilibration at these two energies.
In particular, on the left side of the origin the N/Z ratio of the residue
is more affected by that of the target while on the right side it is more
affected by that of the projectile. However, it should be stressed that the
N/Z ratios on the two sides are not simply those of the target and projectile
but a combination of the two depending on the complicated reaction dynamics.
In the case of $E_{beam}/A=$ 55 MeV, at the time of about 200 fm/c
the heavy residue has broken up into two pieces with some longitudinal
collectivity. The forward moving residue has
an average N/Z ratio of about 7.2/6.0 and an excitation energy of
about 8.6 MeV/nucleon. While the backward moving residue has an
average N/Z ratio of about 21/18 and an excitation energy of
about 6.8 MeV/nucleon.

The proper decay of these primary hot residues is beyond the scope of
the BUU model. Nevertheless, one can imagine that the decay of these
forward-backward moving hot residues with the different N/Z ratios
formed in reactions at higher energies
will result in a strong isospin asymmetry in space. It is also quite
understandable that the isotopic composition of fragments emitted from these
residues will have much greater dependence on the N/Z ratios of the target and
projectile, but not much on the $(N/Z)_{cs}$ of the initial composite systems.
Indeed, this expectation is further evidenced in our systematic studies
on the system $^{40}S+^{58}Zn$, $^{40}Ca+^{58}Fe$ and $^{40}Ti+^{58}Cr$ at
the same beam energies. These features are in good agreement with the
indication of the experimental data at $E_{beam}/A=$ 53 MeV.
Furthermore, the properties of the primary residues
calculated in the above are useful as inputs for a hybrid model
study to compare with the data more quantitatively. Although
there is no global isospin equilibration in reactions at higher
energies, it may not be such a bad approximation to assume the
establishment of isospin equilibrium in the two primary residues formed
at $E_{beam}/A=$ 55 MeV.
One can thus couple the BUU model with statistical models
to compare more quantitatively with the experimental
data. These studies are now in progress and results will be published
elsewhere.

In summary, we have studied the question of
isospin equilibration in
intermediate energy heavy-ion collisions using the isospin-dependent
BUU model. We found that there exists a transition from the isospin
equilibration to non-equilibration as the beam energy increases from
below to above the Fermi energy in central, asymmetric heavy-ion collisions.
At beam energies around 55 MeV/nucleon, the composite system in isospin
non-equilibrium breaks up into two forward-backward moving hot
residues with N/Z ratios more closely related to those of the target and
projectile on the time scale of thermal
equilibrium. These features are in good agreement with that found in
recent experiments at NSCL/MSU and TAMU.

We would like to thank W. Bauer, C. M. Ko and J. Randrup for helpful
discussions. We are also grateful to W. U. Schr\"oder and L.G. Sobotak for
communications on the subject of this study.
This work was supported in part by the NSF Grant No. PHY-9212209,
DOE Grant FG05-86ER40256 and the Robert A Welch Foundation under
Grants A-1110 and A-1266. One of us (SJY) also acknowledge support
from an NSF National Young Investigator Award.

\section*{Figure Captions}
\begin{description}
\item{ Fig.\ 1} \ \ \
Density contours in the reaction plane
at t=200 fm/c for head-on collisions of
Ar+Ni at $E_{beam}/A=$25, 35, 45 and 55 MeV, respectively.
The solid contours are for $\rho=\rho_0/8$, while the dotted
lines are for $\rho=0.05\rho_0$.
\item{ Fig.\ 2} \ \ \
The evolution of the quadrupole moment per nucleon of heavy residues
formed in the head-on collisions of Ar+Ni at $E_{beam}/A=$ 45 and 55 MeV.
\item{ Fig.\ 3} \ \ \
The numbers of neutrons and protons in the residue with
$\rho\geq \rho_0/8$ on the left ($Z < 0$) and right ($Z\geq 0$) side of
the origin in the head-on reactions of Ar+Ni at $E_{beam}/A=$25, 35,
45 and 55 MeV respectively. The solid lines are the numbers of protons
from the projectile, while the dotdashed lines are the numbers of
neutrons from the projectile. The dashed lines are the numbers of protons from
the target, while the dotted lines are the numbers of neutrons from
the target.
\end{description}
\end{document}